\documentstyle[prl,aps,floats]{revtex}

\input epsf.sty

\def\prb{Phys. Rev. B}
\def\prl{Phys. Rev. Lett.}

\def\be{\begin{equation}}
\def\ee{\end{equation}}
\def\ba{\begin{eqnarray}}
\def\ea{\end{eqnarray}}

\def\YBCO{YBa$_2$Cu$_3$O$_{7-\delta}$}

\def\C60{A$_x$C$_{60}$}

\def\hts{high temperature superconductors}

\begin{document}

\twocolumn[\hsize\textwidth\columnwidth\hsize\csname@twocolumnfalse\endcsname

\title
{Classical Phase Fluctuations in High Temperature Superconductors}

\author{E.~W.~Carlson$^{1}$, S.~A.~Kivelson$^{1}$, V.~J.~Emery$^{2}$, and 
E.~Manousakis$^{3}$}
\address
{1)  Dept. of Physics,
U.C.L.A.,
Los Angeles, CA  90095,
2)  Dept. of Physics,
Brookhaven National Laboratory,
Upton, NY  11973-5000,
3)  Department of Physics and MARTECH, Florida State
University, Tallahassee, FL 32306-4350.}
\date{\today}
\maketitle

\begin{abstract}

Phase fluctuations of the superconducting order parameter play a 
larger role in the cuprates than in conventional BCS 
superconductors because of the low superfluid density $\rho_s$ of a
doped insulator. In this paper, we analyze an 
$XY$ model of classical phase fluctuations in the high temperature 
superconductors using a low-temperature expansion and Monte 
Carlo simulations.  In agreement with experiment, the value
of $\rho_s$ at temperature $T=0$ is a quite robust predictor of $T_{c}$,
and the evolution of $\rho_s$ with $T$, including 
its $T$-linear behavior at low temperature, is insensitive to microscopic
details.

\

\end{abstract}

]

Two classes of thermal excitations are responsible for 
disordering the ground state of a superconductor: fluctuations of the
amplitude and phase of the complex order parameter.  
A consensus has not yet been reached on the relative importance of the two
in the {\hts}, since both are anomalous.
The low superfluid density (phase stiffness) of the doped insulator
implies that phase fluctuations 
play an unusually large role.\cite{nature}
Yet the nodes in the $d$-wave gap function support more quasiparticle
(amplitude) excitations at low temperatures than in a {\it clean}
s-wave superconductor.

This paper is concerned with an analytical and numerical study
of the thermal evolution of the in-plane helicity modulus,
$\gamma_{\parallel}(T)$, of an 
anisotropic quasi two-dimensional classical XY model 
of phase fluctuations in a high temperature superconductor.\cite{hm}
We neglect quasiparticle fluctuations because the nodal
quasiparticles are excitations of the insulating state;
very little charge transport is associated with them, and
their contribution
to the superfluid density should be proportional to a positive
power of the (small) doping concentration.\cite{rome2}  We also neglect collective 
amplitude fluctuations associated
with the quantum dynamics of the phase since, with sufficient screening, 
the phase fluctuations are predominantly classical 
down to quite low temperature.\cite{badmetal} 

The calculations focus on the scaled curve,
$\gamma_{\parallel}(T)/\gamma_{\parallel}(0)$ 
vs. $T/T_{c}$, and the value of the 
dimensionless ratios $A_{1}=T_{c}/\gamma_{\parallel}(0)$ and
$A_{2}=T_{c}\gamma_{\parallel}^{\prime}(0)/
\gamma_{\parallel}(0)$. 
(Here $\gamma_{\parallel}^{\prime}(T) \equiv d\gamma_{\parallel}(T)/dT$.)
These nonuniversal quantities turn out to be
rather insensitive to microscopic details of the model, such as
the strength of the interplane coupling and the exact short-distance
nature of the interactions, as shown in Fig. 1 and in the tables.
Figure 1 also shows that the model results agree well with 
experiment,\cite{hardy} when
the helicity modulus of the model is related to the in-plane superfluid 
density, $\rho_{s}$ as determined by
\be 
{\gamma_{\parallel}(T)\over a_{\perp}} = 
{\hbar^{2} {\rho_s(T)  }\over 4 m^*}=
{(\hbar c)^2 \over 16 \pi e^2 \lambda_{ab}^2(T)},
\ee
where $a_{\perp}$ is the spacing 
between planes and $\lambda_{ab}$ is the London penetration depth
within the CuO$_2$ planes.

\begin{figure}[htb]
\centerline{\epsfxsize=3.3in \epsffile{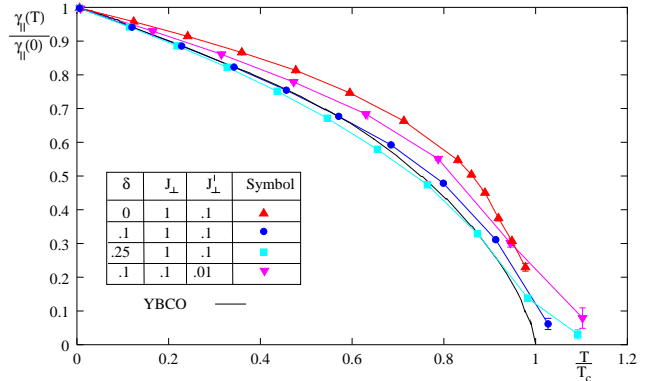}}
\caption{Superfluid density {\it vs.} temperature, scaled
by the zero-temperature superfluid density and by $T_c$, respectively.
Experimental data on $YBCO$ is depicted by the black line, and
is taken from Hardy {\it et al}.\protect \cite{hardy}  
(The data are essentially the same for a range of doping
concentration.)  Our Monte Carlo results 
for system size $10\times10\times10$ are the unfilled symbols.
Calculations are for two planes per unit cell, with coupling
$J_{\parallel}=1$ within each plane, and $J_{\perp}$ and $J_{\perp}^{\prime}$ between alternate planes, as defined in Eq. (3).  
Monte Carlo points above $T_{c}$ are nonzero due to finite size effects.}

\label{fig1}
\end{figure}

There is strong empirical evidence 
that classical phase fluctuations determine 
much of the important physics in the superconducting state of 
the high $T_{c}$ superconductors, and also some properties
of the normal state, especially in underdoped 
materials.\cite{nature}  Most notably, $T_{c}$ increases roughly 
linearly with the zero temperature superfluid 
density,\cite{uemura} ($A_{1}=T_{c}/\gamma_{\parallel}(0)
\sim 1$) 
whereas the 
characteristic energy scale for pairing, $\Delta_o/2$, is both
quantitatively large compared to $k_{B}T_{c}$ and decreases as the doping
increases. Furthermore, ARPES and other measurements of the 
superconducting gap reveal
that pair formation occurs at a crossover temperature well
above $T_{c}$.\cite{shen,levi,doniach}
It is important to note that 
$A_{2}=T_{c}\gamma_{\parallel}^{\prime}(0)/ \gamma_{\parallel}(0)$ 
is roughly constant for various materials and doping concentrations.
This implies that the fluctuations predominantly responsible for the $T$-linear
dependence of the superfluid density at low $T$ are also responsible for
the ultimate destruction of the superconducting state at $T_{c}$.
As first 
pointed out by Roddick and Stroud\cite{stroud}, this behavior is 
characteristic of classical phase fluctuations.

The following arguments have been made
against this interpretation of the data:  1) 
The {\it non-universal} ratio
$A_{1}=T_{c}/\gamma_{\parallel}(0)$ should \cite{gil} theoretically lie in the range 4-8, 
rather than in the                          
experimentally observed range of 0.5 - 1.
In particular, it has been argued
that $A_{1}\propto n$ in multilayer materials, 
where $n$ is the number of layers per unit cell.  2)  
For weakly coupled layers, a phase only model
would \cite{il} yield a $\rho_{s}(T)$ curve that looks like a rounded
Berezinskii-Kosterlitz-Thouless ($BKT$) discontinuity,
unlike what is seen in experiments. 
3)  
Quantum effects suppress classical phase fluctuations \cite{millis} 
for temperatures below the plasma frequency.
4)  If pairing occurs in a substantial range
of temperatures above $T_{c}$, the effects of 
fluctuation superconductivity should be observed, contrary to 
experiment.\cite{gil}  

As can be seen from the figure and the tables, our 
present results conclusively show that the first two assertions 
are incorrect;  
the classical XY model is quantitatively consistent 
with experiments.  
The third suggestion has been previously shown to be incorrect\cite{badmetal}, 
due to screening by the substantial background normal fluid.  
Specifically, in a two-fluid model of a superconductor, the classical 
model is reliable down to a classical to quantum crossover temperature
which can be well below $T_{c}$:  $T_{class} \propto T_{c}/\sigma_{N}$,
where $\sigma_{N}$ is an 
average of the optical conductivity of the normal component in units 
of the quantum of conductance.
The fourth point overlooks the fact that fluctuation 
superconductivity is only significant close to T$_c$ where the
correlation length is long.  In
conventional superconductors the observed fluctuations
involve amplitude and phase and are Gaussian,
while for the high $T_{c}$ superconductors, 
true critical fluctuations in the XY universality class are
detected in a remarkably broad range of temperatures.\cite{3dxy} 

A classical XY model on a tetragonal lattice will be used to
study  the effects of phase fluctuations in a 
quasi two-dimensional superconductor 
at wavelengths that are long enough for amplitude fluctuations to be unimportant.
The in-plane unit cell area $a_{\parallel}^{2}$ does not enter into the evaluation
of T$_c$ or the temperature dependence of $\gamma_{\parallel}(T)$. Here
$a _{\parallel}$ is a short-distance cutoff
which will be discussed at the end of the paper. 
In general, the interaction
energy, $V$, depends on the phase difference, $\theta_{ij}\equiv
\theta_{i}-\theta_{j}$, between nearest-neighbor sites $<i,j>$.
Because of gauge invariance and time reversal symmetry, $V$ can
be expanded in a cosine series,
\be
V(\theta_{ij}) = \sum_{n} A_n \cos(n \theta_{ij}).
\label{eq:V}
\ee
The first harmonic, $\cos(\theta)$, corresponds to the transfer of one 
pair of electrons
between neighboring cells;  
each successive harmonic 
transfers a higher number of pairs.  
We keep only the first two terms in the cosine series for couplings 
within 
a plane, and the first cosine term
for the weaker Josephson coupling between planes.

\ba
H= && -J_{\parallel} \sum_{<ij>_{\parallel}} \left\{ 
\cos(\theta_{ij}) 
+ \delta\cos(2\theta_{ij})\right \} \nonumber \\ 
- &&  \sum_{<kl>_{\perp}} \left\{
J_{\perp}^{kl}\cos(\theta_{kl}) \right \},
\label{eq:model}
\ea
where 
the first sum is over nearest neighbor sites within each plane, 
and the second sum is over nearest neighboring
planes.  The coupling, $J_{\parallel}$, will be assumed 
to be isotropic within each plane and
the same for every plane, but the coupling between planes,
$J_{\perp}^{kl}$, is different for crystallographically
distinct pairs of neighboring planes.  
It will be assumed that $J_{\parallel}$, $J_{\perp}$, and $\delta$ are 
positive, since there is no reason to expect any frustration
in the problem,\cite{spivak}
and that $\delta \le 0.25$, since for $\delta > 0.25$ there is 
a secondary minimum in the potential for $\theta_{ij} = \pi$,
which is probably unphysical. 
The sensitivity of various computed 
quantities to variations in $\delta$ in this range  is a measure 
of the importance of ``microscopic details.''
                                                                                                                              
It follows from simple and general considerations
that most features of the thermal evolution of the
superfluid density of {\YBCO} shown in Fig. 1 are reproduced by such a 
model. The critical phenomena 
are in the same universality class as the classical 3D XY 
model, which is consistent with the observed behavior\cite{3dxy} 
in {\YBCO} within about 10\% 
of T$_{c}$.  Furthermore, the helicity modulus is
linear in the temperature, as observed in {\YBCO}.  Indeed, 
using linear spin-wave theory, it is straightforward to obtain 
the first terms in the 
low-temperature series for $\gamma$:
\ba
{\gamma_{\parallel}}(T) = J_{\parallel}(1 + 4 \delta) - {\alpha
(1 + 16 \delta) \over 4 
(1 + 4 \delta)}T + {\cal O}(T^2),
\label{eq:rho2}
\ea
where $\alpha$
is a numerical integral which
varies from $\alpha=1$ in the 2d limit
 ($J_{\perp} \rightarrow 0$), to $\alpha=2/3$ for 
$J_{\perp} \rightarrow \gamma_{\parallel}(0)=
J_{\parallel}(1 + 4\delta)$ for one plane per unit cell.

A more quantitative comparison between the classical model and 
experimental data can be undertaken by studying various dimensionless
ratios, particularly 
$A_{1}=T_{c}/\gamma_{\parallel}(0)$ and
$A_{2}=T_{c}\gamma_{\parallel}^{\prime}(0)/
\gamma_{\parallel}(0)$.
Here, $T_{c}$ is computed numerically by means of the Binder parameter\cite{binder}
for systems of size up to $24\times24\times24$.
Errors in $T_{c}$ are limited by the resolution with which 
the Binder crossing point is computed in each case.
The quantities $\gamma_{\parallel}(0)$ and 
$\gamma_{\parallel}^{\prime}(0)$
are obtained from Eq. (\ref{eq:rho2}).

Experimentally $A_2 \sim \frac 1 2$, and $A_1$ is in the range
0.6-1.3 for underdoped and optimally doped materials.
Tables 1 and 2 show the ratios $A_{1}$ and $A_{2}$ for 
various choices of parameters in the classical XY model.
Note that
$A_2 \sim \frac 1 2$  for $\delta$ not too small, 
whereas
$A_2$ is about a factor of two smaller for $\delta=0$.
The shape of the $\rho_{s}(T)$ vs. $T$ curves, as quantified
by $A_2$, is remarkably robust, especially if we compare the cases
of $\delta=0.1$ to $\delta=0.25$.
The ratio $A_1$ is  a little more sensitive
to the value of $\delta$, but it is comfortably in the experimental range
for $\delta$ between 0.1 and 0.25, and only slightly larger for $\delta=0$.
The relative insensitivity of $A_{1}$ to $\delta$ and to 
the details of the interplane couplings, $J_{\perp}$, 
demonstrates that when classical phase fluctuations govern the 
physics, $\rho_{s}(0)$ is a quantitatively good predictor of $T_{c}$.

\begin{center}
\begin{tabular}{|c||c|c|c||c|c|c||c|c|c|}
\hline
{$J_{\perp}$}
& {0} & {0} & {0} & {0.01}& {0.01}& {0.01} & {0.1} & {0.1} & {0.1}\\
\hline
{$\delta$}
& {0} & {0.1} & {0.25} & {0} & {0.1} & {0.25} & {0} & {0.1} & {0.25} 
\\
\hline
{$A_1$}
& {0.89} & {0.72} & {0.6} & {1.1}  & {.828}  & {.625}  & {1.324}  & 
{.986}  & {.73}\\
\hline
{$A_2$} 
& {.22} & {.335} & {.38} & {.27} & {.381} & {.388} & {.3066} & {.432} 
& {.437}\\
\hline
{$T_{c}$}
& {.89} & {1.01} & {1.2} & {1.1} & {1.16} & {1.25} & {1.324} & {1.38} 
& {1.46} \\
\hline
\end{tabular}
\end{center}

Table 1: {\it Single Layer}
The dimensionless ratios $A_1 = T_{c}/\gamma_{\parallel}(0)$ and 
$A_{2}=T_{c}\gamma_{\parallel}^{\prime}(0)/
\gamma_{\parallel}(0)$
which characterize the superfluid density vs. temperature
for systems with one layer per unit cell.  $J_{\perp}$ 
and $T_{c}$ are
quoted in units of $J_{\parallel}$.

\

\begin{center}
\begin{tabular}{|c||c|c|c||c|c|c|}
\hline
{$J_{\perp}$} & {1 : 0} & {1 : 0} & {1 : 0} & {.1 : 0} & {.1 : 0} & 
{.1 : 0} \\ 
\hline
{$\delta$} & {0} & {0.1} & {0.25} & {0} & {0.1} & {0.25} \\
\hline
{$A_1$} & {1.38} & {1.03} & {0.78} & {1.13} & {0.836} & 
{0.645} \\
\hline
{$A_2$} & {.279} & {.402} & {.426} & {.271} & {.376} & 
{.394} \\
\hline
\hline
\hline
{$J_{\perp}$} & {1 : .01} & {1 : .01} & {1 : .01} & {.1 : .01} & {.1 
: .01} & {.1 : .01} \\
\hline
{$\delta$} & {0} & {0.1} & {0.25} & {0} & {0.1} & {0.25} \\
\hline
{$A_1$} & {1.52}  & {1.12}  & {0.83}  & {1.2}  & {0.907}  & 
{0.675} \\
\hline
{$A_2$} & {.306} & {.437} & {.452} & {.29} & {.407} & {.411} 
\\
\hline
\hline
\hline
{$J_{\perp}$} & {1 : .1} & {1 : .1} & {1 : .1} & {.1 : .1} & {.1 : 
.1} & {.1 : .1} \\
\hline
{$\delta$} & {0} & {0.1} & {0.25} & {0} & {0.1} & {0.25} \\
\hline
{$A_1$} & {1.6975}  & {1.252}  & {0.916} & {1.324}  & 
{0.986}  & {0.73}\\
\hline
{$A_2$} & {.3318} & {.4772} &{.4906} & {.3066} & {.432} & 
{.437} \\
\hline
\end{tabular}
\end{center}

Table 2:  {\it Double Layer}
The dimensionless ratios $A_1 = T_{c}/\gamma_{\parallel}(0)$ and
$A_{2}=T_{c}\gamma^{\prime}_{\parallel}(0)/
\gamma_{\parallel}(0)$,
which characterize the superfluid density vs. temperature
curve, for systems with two layers per unit cell;  the two 
values of $J_{\perp}$ are between planes within a unit cell and
between unit cells.

\

Hardy {\it et al.} have demonstrated\cite{hardy} that 
when $\rho_{s}(T)/\rho_{s}(0)$ is plotted {\it vs.} $T/T_{c}$, for
various dopant concentrations in {\YBCO}, the data collapse 
approximately onto one curve.  Since $T_{c} \propto \rho_{s}(0)$, this
amounts to scaling out $T_{c}$ for both axes.  Thus the unique shape of 
the normalized $\rho_{s}$ vs. $T$ curve may be attributed to
the existence of a single energy scale, the transition temperature. 
This is inconsistent with scenarios that require the superfluid density 
to be depleted by quasiparticle excitations as the temperature is raised,
since the energy scale is set by $\Delta_{0}$, and
$\Delta_{0}/T_{c}$ 
is not roughly constant, but varies by a factor of 2
in the range of doping studied by Hardy {\it et al}.  

We have also used Monte Carlo calculations to evaluate the
superfluid density for $0 < T < T_{c}$.
The results are scaled as in Hardy {\it et al.}\cite{hardy} and
compared with their data in Fig. 1.  
As anticipated, for $\delta$ not too near zero, the model 
is insensitive to the
value of $\delta$, and in remarkable agreement with
experiment, considering that no parameters have been tuned.  
Quasiparticle excitations near the nodes of the $d$-wave
gap may give a $T$-linear contribution to the superfluid density,
but our results suggest 
that this contribution is quantitatively small.  This, in turn,
supports the idea that there is little charge transport
associated with the quasiparticles.

A much-discussed feature of the systematics of $T_c$ in the {\hts} \cite{sudip}
is the observed increase of $T_{c}$ within each family of materials with 
the number of planes per unit cell, $n$.  Within the classical phase 
model, the fact that phase fluctuations lead to a particularly large 
suppression of $T_{c}$ below its mean-field value (see Table 3) 
leads to an increased 
sensitivity to even weak couplings in the third direction.
This produces a strong increase of $T_{c}$ with $n$, although
possibly not quite as strong as observed experimentally.
However, it should be noted that other things may change with $n$; for
example, in a three-plane material, the central plane may have a 
different hole concentration than the others. 
At the same time, since the dimensionless constant 
$\alpha$ in Eq. (\ref{eq:rho2}) is
only weakly dependent on 
$J_{\perp}$, $A_2$ is an increasing function of $n$.  
This is to be expected as small values of $A_2$ are
characteristic of 
low dimensional phase fluctuations.
Table 3 shows the variation of $T_c$ with number of planes
per unit cell for weak coupling between planes.

\begin{center}
\begin{tabular}{|c|c|c|c|c|c|}
\hline
{$\delta$}
& {0} & {0} & {0} & {0} & {0} \\
\hline
{$n$} 
& {1} & {2} & {3} & {4} & {$\infty$} \\
\hline
{${d\rho(0) \over dT}$}
& {.2472} & {.2384} & {.2365} & {.2348} & {.2315}\\
\hline
{$T_{c}$}
& {1.09} & {1.20} & {1.24} & {1.26} & {1.324} \\
\hline
{$T_{MF}$}
& {1.111} & {1.287} & {1.334} & {1.361} & {1.394} \\
\hline
\hline
\hline
{$\delta$}
& {0.1} & {0.1} & {0.1} & {0.1} & {0.1} \\
\hline
{$n$} 
& {1} & {2} & {3} & {4} & {$\infty$} \\
\hline
{${d\rho(0) \over dT}$}
& {.4604} & {.4486} & {.4450} & {.4427} & {.4379} \\
\hline
{$T_{c}$}
& {1.16} & {1.27} & {1.30} & {1.32} & {1.38} \\
\hline
\hline
\hline
{$\delta$}
& {0.25} & {0.25} & {0.25} &  {0.25} & {0.25} \\
\hline
{$n$} 
& {1} & {2} & {3} & {4} & {$\infty$} \\
\hline
{${d\rho(0) \over dT}$}
& {.6211} & {.6092} & {.6055} & {.6032} & {.5982} \\
\hline
{$T_{c}$}
& {1.25} & {1.35} & {1.39} & {1.41} & {1.46} \\
\hline
\end{tabular}
\end{center}

Table 3:  {\it Variations as a function of
the number of planes per unit cell:}
The coupling between planes within
the unit cell is $J_{\perp} = 0.1$, and 
between planes of different unit cells is $J_{\perp}^{\prime} = 0.01$.
$T_{MF}$ is the interplane
mean-field estimate of $T_{c}$ obtained as described in the text.

\

Mean field theory is a standard method of estimating 
the effects of weak higher-dimensional 
couplings on the critical temperature of quasi one or two 
dimensional systems. 
For instance, for one plane per unit cell ($n=1$)
this approach leads 
to an implicit equation for the three dimensional $T_{c}$
\be
\chi_{2d}(T_{MF})2J_{\perp}=1,
\ee
where $\chi_{2d}(T)$ is the 
susceptibility of an isolated plane.  For the case 
$\delta=0$, we have computed the interplane mean-field transition 
temperature, which is also presented in Table 3, using the Monte Carlo 
results of Gupta and Baillie\cite{gupta} for $\chi_{2d}(T)$.  This 
mean-field theory becomes exact in the limit 
$J_{\perp}\rightarrow 0$, and always gives an upper bound to $T_{c}$.  

Phase fluctuations should also have detectable effects on 
other equilibrium properties, such as the specific heat, 
the diamagnetic susceptibility, and $\gamma_{\perp}$.  
In contrast to $\gamma_{\parallel}$, these quantities
depend on $a_{\parallel}$.
Classical systems have nonzero specific heat at $T=0$, in violation
of the third law of thermodynamics.
The classical XY model at temperatures $T \ll T_{c}$ has a specific 
heat per unit area in a CuO$_2$ plane equal to $C= k_{B}/2 a_{\parallel}^{2}$. 
The specific heat\cite{specific} 
at $T=2K$ of good crystals of optimally doped 
YBCO is roughly $5\times 10^{-4}k_{B}$ per planar copper;  
if we assumed that all of 
this specific heat were due to phase fluctuations, it would imply
a$_{\parallel}=32$ lattice constants.  

In the classical XY model
a$_{\parallel} \sim$ r$_v$, where r$_v$ is the radius of a vortex core.
Recent $\mu$SR measurements\cite{musr}
have found that r$_v$ grows substantially at low fields, and tends to 
a zero field value which is on the order of 100$\AA$ (26 lattice 
constants) and which is only weakly temperature dependent nearly up 
to $T_{c}$.  This behavior is consistent with other measurements, especially 
STM in a field of 6T,\cite{STM} which have
produced significantly smaller estimates of the core radius.
Thus, if we estimate $a_{\parallel}$
using the $\mu$SR data, it is consistent to attribute a large fraction of the 
low-temperature specific heat to classical phase fluctuations. 
The contribution of critical fluctuations to the specific heat near 
$T_{c}$ may also be dominated by classical phase fluctuations, but a 
quantitative comparison of the theoretically expected (non-universal) 
critical amplitudes with experiment is not straightforward.

Experimentally, $\rho_{\perp}/m^{*}$ is much more weakly temperature 
dependent than $\rho_{\parallel}/m^{*}$ at low temperatures.  We 
intend in the near future to analyze the implications of this 
observation for the phase fluctuation model.  Specifically, since
$\gamma_{\perp}/\gamma_{\parallel}=
\rho_{\perp}/\rho_{\parallel}(a_{\parallel}/a_{\perp})^{2}$,
the anisotropy of the superfluid density may also 
contain valuable information concerning the physics of $a_{\parallel}$.

Finally, we address the remarkable measurements of the frequency 
dependent superfluid density in BSCCO of Corson {\it et 
al}.\cite{corson}  
Without making any explicit assumptions 
concerning the dynamics, we can interpret these results in terms of a 
finite size scaling hypothesis, in which we associate a length scale,
$L(\omega)$, with the finite measurement frequency, and
\be
\gamma(T,L)\sim L^{x}T\tilde \gamma(L/\xi(T))
\ee
where $\xi(T)$ is the correlation length of the infinite system at 
temperature $T$.  Since BSCCO is highly anisotropic, we follow
Corson {\it et al}\cite{corson}  in assuming that the finite 
frequency response 
is essentially two dimensional, in which case $x=0$, and
$\xi(T)$ is infinite for all $T<T_{BKT}$, the BKT 
transition temperature.  This implies that $\gamma$ is approximately
frequency independent for $T<T_{BKT}$ and vanishes exponentially as
a function of $L/\xi$ at temperatures enough above $T_{KT}$ that
$L>\xi(T)$.  Indeed, $\gamma(T,L)$ was computed numerically
for the two dimensional
XY model by Schultka and Manousakis\cite{sm}
and we have repeated these calculations for anisotropic three dimensional 
models; the results confirm 
that our model nicely accounts for the
observations of  Corson {\it et 
al}.\cite{corson}, with the proviso that the measured
finite frequency superfluid density is interpreted as a renormalized, 
rather than a ``bare'', response function.

{\bf Acknowledgements:}  We acknowlege stimulating 
discussions with D-H.Lee, E. Fradkin,
K. Moler, A. Millis, L. Ioffe, and K. Nho.  This work was 
supported in part by NSF grant number DMR98-08685 
(EWC and SAK), 
DOE grant number DE-AC02-98CH10886 (VJE), and by NASA grant number NAG3-1841 (EM).

\end{document}